\documentclass[letter,usenatbib]{mnras}

\usepackage{newtxtext,newtxmath}
\usepackage[T1]{fontenc}

\DeclareRobustCommand{\VAN}[3]{#2}
\let\VANthebibliography\thebibliography
\def\thebibliography{\DeclareRobustCommand{\VAN}[3]{##3}\VANthebibliography}

\usepackage{graphicx}
\usepackage{dcolumn}
\usepackage{amsmath}    
\usepackage[para]{threeparttable}
\usepackage{verbatim}   
\usepackage{subfigure}  
\usepackage{gensymb} 
\usepackage{wasysym} 
\usepackage{hyperref}   
\usepackage{nth}
\newcommand{\kms}{\mbox{${\rm km\,s}^{-1}$}}
\newcommand{\ms}{\mbox{${\rm m\,s}^{-1}$}}

\newcommand{\Planet}{WASP-166~b}
\newcommand{\Rjup}{\mbox{${R}_{J}$}}

\title[Confirmation of sodium in \Planet{}]{The Hot Neptune \Planet{} with ESPRESSO II: Confirmation of atmospheric sodium\thanks{Based on observations made at ESO's VLT (ESO Paranal Observatory, Chile) under ESO programme 106.21EM (PI Cegla).}}
\author[J.~V.~Seidel et al.]{J.~V.~Seidel,$^{1}$\thanks{ESO Fellow, jseidel@eso.org}
H.~M.~Cegla,$^{2,3}$\thanks{UKRI Future Leaders Fellow}
L.~Doyle,$^{2,3}$
M.~Lafarga,$^{2,3}$
M.~Brogi,$^{2,3,4}$
S.~Gandhi,$^{2,3,5}$ 
\newauthor
D.~R.~Anderson,$^{2,3}$
R.~Allart,$^{6}$
N.~ Buchschacher,$^{7}$ 
C.~Lovis,$^{7}$ 
and D.~Sosnowska$^{7}$ 
\\
$^{1}$European Southern Observatory, Alonso de C\'ordova 3107, Vitacura, Regi\'on Metropolitana, Chile\\
$^{2}$Centre for Exoplanets and Habitability, University of Warwick, Coventry, CV4 7AL, UK\\
$^{3}$Department of Physics, University of Warwick, Coventry, CV4 7AL, UK\\
$^{4}$Osservatorio Astrofisico di Torino, Via Osservatorio 20, 10025, Pino Torinese, Italy\\
$^{5}$Leiden Observatory, Leiden University, Postbus 9513, 2300 RA Leiden, The Netherlands\\
$^{6}$Department of Physics, and Institute for Research on Exoplanets, Universit\'e de Montr\'eal, Montr\'eal, H3T 1J4, Canada \\
$^{7}$Observatoire Astronomique de l'Universit\'e de Gen\`eve, Chemin Pegasi 51b, CH-1290 Versoix, Switzerland
}

\date{Accepted XXX. Received YYY; in original form ZZZ}

\pubyear{2022}

\begin{document}
\label{firstpage}
\pagerange{\pageref{firstpage}--\pageref{lastpage}}
\maketitle

\begin{abstract}
The hot Neptune desert, a distinct lack of highly irradiated planets in the size range of Neptune, remains one of the most intriguing results of exoplanet population studies. A deeper understanding of the atmosphere of exoplanets sitting at the edge or even within the Neptune desert will allow us to better understand if planetary formation or evolution processes are at the origin of the desert. A detection of sodium in \Planet{} was presented previously with tentative line broadening at the $3.4\,\sigma$ with the HARPS spectrograph. We update this result with two transits observed with the ESPRESSO spectrograph, confirming the detection in each night and the broadened character of the line. This result marks the first confirmed resolved sodium detection within the Neptune desert. In this work, we additionally highlight the importance of treating low-SNR spectral regions particularly where absorption lines of stellar sodium and planetary sodium overlap at mid-transit - an important caveat for future observations of the system.

\end{abstract}

\begin{keywords}
Planetary Systems -- Planets and satellites: atmospheres, individual: WASP-166b -- Techniques: spectroscopic -- Instrumentation: spectrographs -- Methods: observational
\end{keywords}

\section{Introduction}

Exoplanet science has firmly moved from the continuous stream of new detections to the connections between the existence and character of these worlds in exoplanet population studies. One of the earliest and most curious developments in the exoplanet population is the emergence of the so-called Neptune desert, the lack of strongly irradiated Neptune-sized planets in the radius-insolation space \citep{Lecavelier2007, Beauge2013, Mazeh2016}. With the technological advancements of the last decade, it has become clear that this feature of the exoplanet population is not a detection bias \citep{Mazeh2016}, as these planets should, in theory, be observable with various techniques. Instead, the Neptune desert's origin are most likely due to the increased atmospheric escape in these highly irradiated, but comparatively small planets \citep{Owen2018,Owen2019}. This makes planets within the desert, which are barely holding on to their tenuous atmospheres, perfect laboratories to study atmospheric escape in real time.

\noindent Recently, various groups have started to characterise the tenuous atmospheres within or at the edge of the Neptune desert with detections of atmospheric sodium, hydrogen, helium, and water vapour \citep[e.g. ][]{Armstrong2020, Seidel2020c, Dragomir2020, Crossfield2020, Allart2020, Murgas2021, Brande2022}. In this sample of desert inhabitants, \Planet{} remains a rare gem, a planet that has apparently retained its atmosphere which exhibits extreme bloating with a density of $\rho = 0.54\pm0.09\, \mathrm{g}/\mathrm{cm}^3$ \citep{Hellier2019, Bryant2020}.

\noindent While planets at the edge of the Neptune desert are tantalising targets, they remain challenging in an observational context with current technologies due to their size. Compared to Jupiter-sized targets, Neptune-like planets have a significantly lower atmospheric signal-to-noise ratio (SNR) and need on average more transits for similar results. Here, we present the confirmation of the previous tentative sodium detection in \citet{Seidel2020b} in the atmosphere of \Planet{} with the ESPRESSO spectrograph using two transits. First, we describe the observations and applied data reduction (Section \ref{sec:obs}), then, in Section \ref{sec:fulltrans}, we show the techniques applied to create the transmission spectra, highlighting the influence of low-SNR remnants, and the calculation of the false positive detection probability. Lastly, we discuss the detection of sodium for \Planet{} and the implications for the Neptune desert and future studies (Section \ref{sec:interp}).

\section{Observations and data reduction}
\label{sec:obs}

 \begin{table*}
\caption{Log of ESPRESSO observations.}
\label{table:nightoverview}
\centering
\begin{threeparttable}
\begin{tabular}{c c c c c c c c }
\hline
\hline
&Date   &$\#$Spectra\tnote{$^a$}  &Exp. Time [s]  &Airmass        \tnote{$^b$}&Seeing & SNR \#116 & SNR stellar line core     \\
\hline
Night 1&2020-Dec-31&292 (160/104\tnote{$^c$})&100 &2.6-1.20-1.15&0.6-0.8&30 - 37 & 7-8\\
Night 2&2021-Feb-18&278 (162/114\tnote{$^d$}) &100&1.20-1.10-1.70&0.7-1.3&22 -  35 & 4-8\\
\hline
\end{tabular}
\begin{tablenotes}
\item[$^a$]In-/ out-of-transit spectra after excluding exposures. \item[$^b$]Airmass at the beginning/ centre/ end of transit. \item[$^c$] 13 out-of-transit spectra rejected due to high airmass. \item[$^d$]2 out-of-transit spectra rejected due to telescope vignetting.
\end{tablenotes}
\end{threeparttable}
\end{table*}

We observed the bloated super-Neptune \Planet{} during its transit on two separate nights (2020-Dec-31, 2021-Feb-18) with the ESPRESSO echelle spectrograph installed at ESO's VLT telescopes in Paranal Observatory, Chile \citep{Pepe2021} (ESO programme: 106.21EM, PI: Cegla). \Planet{} orbits a bright F9-type star (Vmag = $9.36$, distance=$113.0\pm1.0$ pc) in $5.44$ days with a radius of $0.63\pm0.03\,\Rjup$ \citep{Hellier2019}.
An overview of the observations can be found in Table \ref{table:nightoverview}.

\noindent This work focuses on the analysis of the planetary sodium doublet (spectral order $116$ and $117$, approximately at $5888 - 5898$~\r{A}), previously detected with the HARPS spectrograph in \citet{Seidel2020b}. The observations span the time where the planet occults part of the stellar disk (in-transit spectra, from T$_1$ to T$_4$) and also before and after the transit (out-of-transit spectra). In night one (see Table \ref{table:nightoverview}), the observations were started at airmass $2.6$, however the ESPRESSO ADC (Atmospheric Dispersion Corrector) is only calibrated for airmass $<2.2$. In consequence the first 13 out-of-transit spectra were rejected from the analysis.The last two spectra in night two were excluded due to telescope vignetting. The obtained spectra are corrected for instrumental effects and sky emissions are subtracted by the ESPRESSO pipeline (version 2.3.1) as implemented in DACE (\url{https://dace.unige.ch}). We monitored telluric sodium emission with the on-sky fibre B and found no  contamination. Cosmic rays were rejected ($5 \sigma$ difference to the mean of all exposures in each wavelength bin) and replaced with the time averaged mean \citep[as in e.g.][]{Wy15, Seidel2019}.

\begin{figure}
\resizebox{\columnwidth}{!}{\includegraphics[trim=2.0cm 2.0cm 2.0cm 1.3cm]{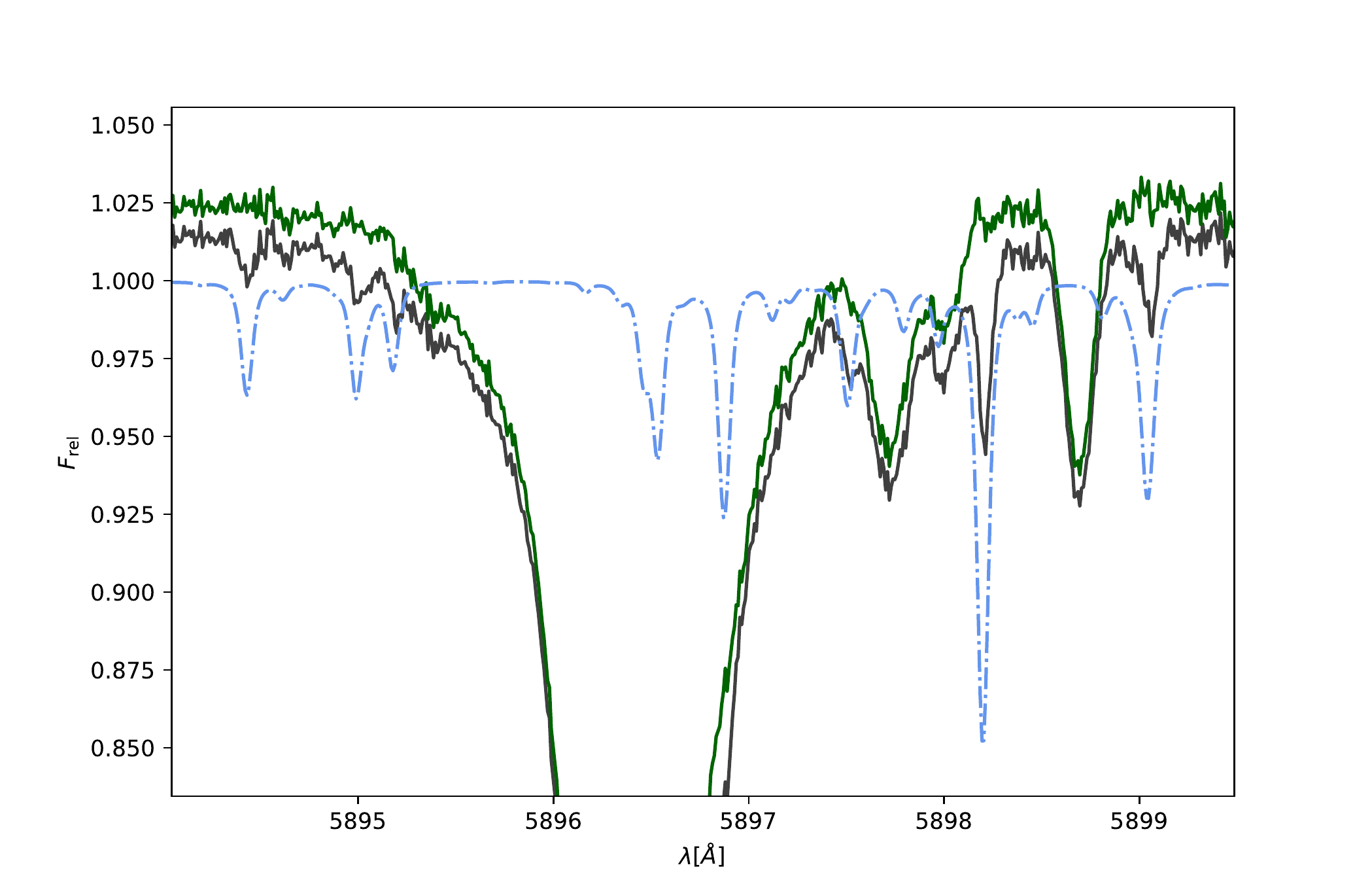}}
        \caption{Telluric correction using {\tt molecfit} in the wavelength range around the $D_1$ sodium line at $~5896\,\AA$ for night two. The master out spectrum before telluric correction (black) and after telluric correction (green) is shown in the observer's rest frame, where telluric absorption lines are aligned in wavelength. A scaled model of the telluric lines calculated for mid-transit is shown in light blue, dotted-dashed lines.}
        \label{fig:tell_corr}
\end{figure}

During both nights, the different velocities (barycentric velocity, stellar, and planetary) led to significant overlap between telluric $H_2O$ absorption lines and the sodium doublet. We corrected the telluric lines with {\tt molecfit}, version 1.5.1. \citep{Sm15, Ka15}. Parameters were set as described in \citet{Al17}. Because molecfit takes into account the atmospheric conditions in the specific observation night, it corrects microtellurics and also stronger telluric absorption lines. We controlled the telluric correction of {\tt molecfit} by comparing the sum over all spectra (master spectrum) before and after its application in the observer's rest frame (ORF). Figure \ref{fig:tell_corr} shows the master spectrum before and after correction, as well as an unscaled model of the telluric lines to indicate the position of the tellurics for night two. The telluric effects were reduced to the noise level and have no effect on our results. Our analysis uses the system parameters presented in \cite{Hellier2019}.

\section{Transmission spectroscopy of \Planet{}}
\label{sec:fulltrans}

We follow \citet{Tabernero2020,Borsa2021,Seidel2021} for the calculation of the transmission spectrum using ESPRESSO data. All spectra were weighted by their SNR and the errors propagated throughout the analysis. The telluric corrected spectra are shifted from the ORF to the stellar rest frame (SRF). In this rest frame all stellar lines are aligned in time and the planetary signal can be separated from the stellar signal by dividing each spectrum taken in-transit by the normalised sum of all out-of-transit spectra (master-out). The separated planetary spectra are then shifted into the planetary rest frame (PRF) where the normalised sum builds the final transmission spectrum. The respective shifts varied for the barycentric velocity between $21.12 - 20.53\, \kms$ in night one and $2.72 - 2.05\, \kms$ in night two. The system velocity was taken from literature as $23.63\, \kms$ \citep{Hellier2019}. The planet velocity ranges from $-14.98$ to $21.60\, \kms$ in night one and from $-20.67$ to $17.78\,\kms$ in night two, with the stellar velocity between $-1.20$ to $1.73\, \ms$ and $1.66$ to $-1.42\, \ms$ respectively.

\begin{figure*}
\resizebox{\textwidth}{!}{\includegraphics[trim=3.0cm 2.0cm 3.0cm 0.8cm]{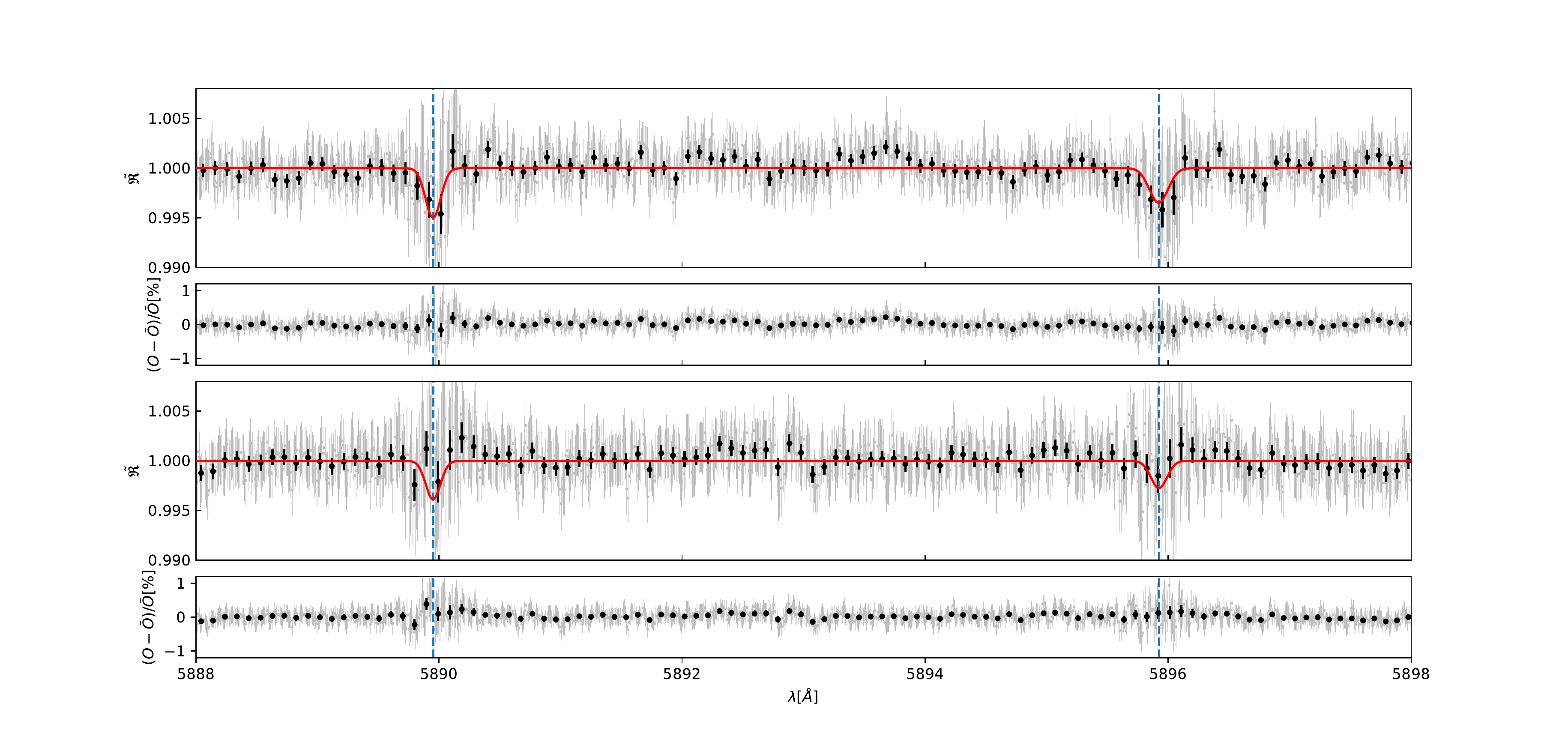}}
        \caption{ESPRESSO sodium doublet for night one (top) and nights 2 (bottom) in the PRF (planetary rest frame). Larger panels: in light gray, the transmission spectrum in full ESPRESSO resolution is shown, in black binned by 20$\times$ for visibility. The theoretical position of the doublet lines are shown as vertical, blue, dashed lines. The Gaussian fits to the unbinned data is shown as a red line. Smaller panels: Residuals of the red Gaussian fit in $\%$. }
        \label{fig:transspectrum}
\end{figure*}

\noindent The Coud\'{e} Train optics of ESPRESSO are known to create interference patterns, which manifest as sinusoidal noise (wiggles) in the transmission spectra of each order \citep{Allart2020,Tabernero2020}. We performed a sinusoidal fit on the transmission spectrum of each order leaving out the sodium doublet lines to obtain a straight continuum. We show both transmission spectra in Figure \ref{fig:transspectrum}) where we combine both spectral orders independently containing the sodium doublet ($116$ and $117$). The Gaussian fit shown in red in Figure \ref{fig:transspectrum} is obtained on the unbinned data, with the depth and width as free fitting parameters.

\subsection{Stellar effects}
\label{sec:stellar}

Two important effects when assessing resolved lines are the Rossiter-McLaughlin (RM)-effect, where the Doppler-shift introduced from the movement of the stellar surface behind the planet influences the planetary detection (for details see \cite{Rossiter1924, McLaughlin1924, Lo15, Cegla2016}) and the center to limb (CLV) variation in the stellar spectra. For \Planet{} the surface radial velocity of the stellar surface occulted by the planet during transit ranges from $\sim -5\,\kms$ to $\sim5\,\kms$. In \citet{Seidel2020b}, the impact of the RM-effect on the transmission spectrum was estimated to be below $0.5\,\%$ at line centre with the conclusion that the RM-effect cannot mask the sodium feature for this particular planet. We applied the same numerical correction of the RM and CLV effects that was applied to the HARPS dataset in \citet{Seidel2020b}, following \citet{Wyttenbach2020} with the local stellar RV model taken from \citet{doyle2022wasp_166}.

\begin{figure}
\resizebox{\columnwidth}{!}{\includegraphics[trim=3.0cm 10.0cm 4.0cm 10.0cm]{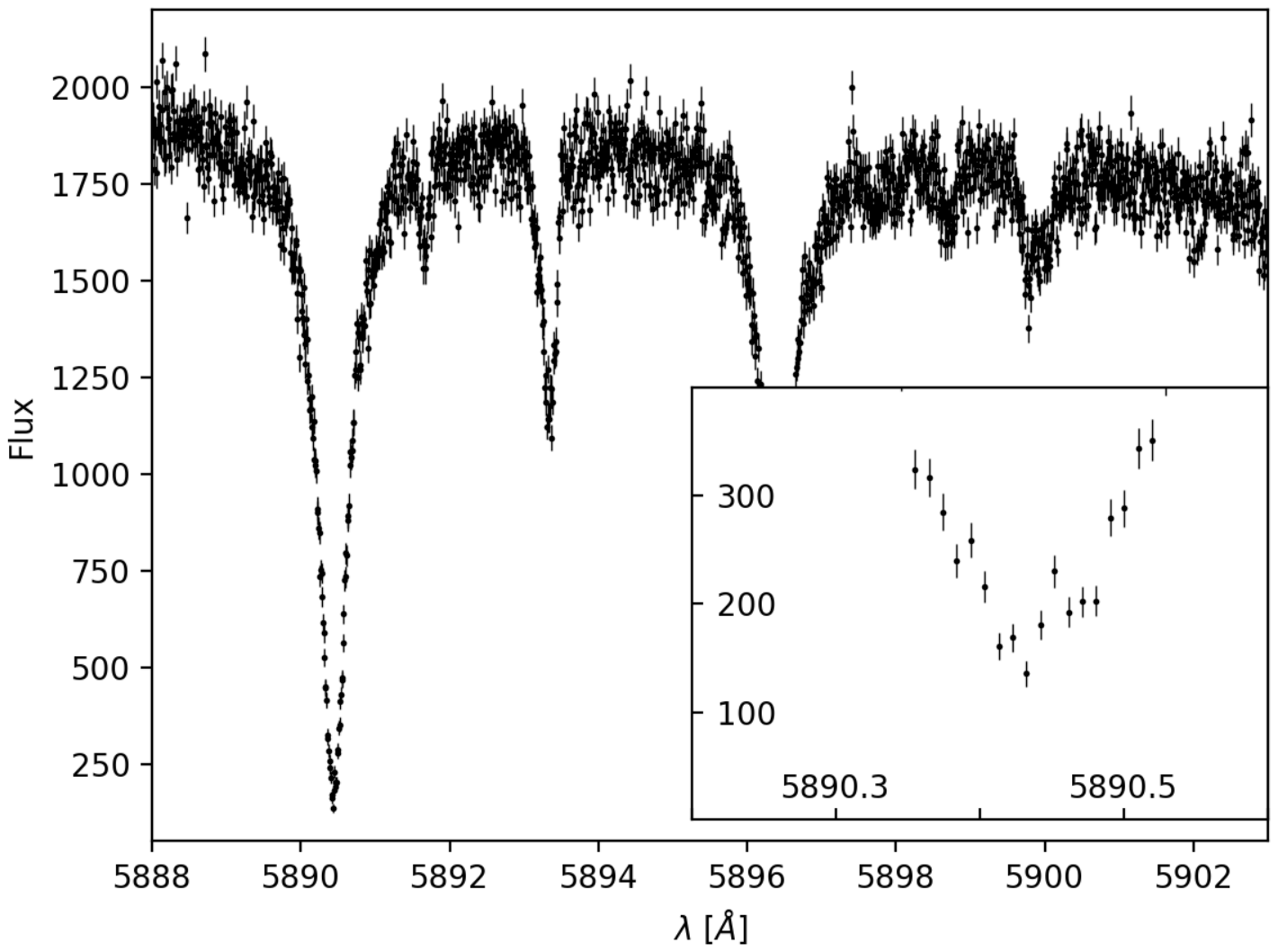}}
        \caption{Spectrum Nr. 105 taken during night two with its original flux level and shape, corrected for telluric absorption lines. The order $116$ is zoomed into the wavelength range of the stellar sodium doublet in the observer's rest frame. The inset shows the centre of the stellar D$_2$ line, highlighting the extremely low stellar flux due to absorption of sodium in the star's photosphere. The masked wavelength range corresponds to the FWHM of the stellar line.}
        \label{fig:low_flux}
\end{figure}

\noindent Due to the different Doppler-shifts coming from the planetary and system movements, the planetary signal moves with respect to the stellar sodium absorption lines. At transit centre, when the planetary absorption is aligned with the stellar absorption, the SNR reduces significantly due to the magnitudes lower flux in the stellar line centre (see Figure \ref{fig:low_flux} and \citet{Barnes2016, Borsa2018,Seidel2020b,Seidel2020c}). For the dataset presented here, this low SNR-remnant is particularly poignant due to the low exposure time (employed to study the system architecture, the main goal of the observing proposal, see \citet{doyle2022wasp_166}) in comparison to the HARPS dataset. For the overall flux level of approximately $100$ photons in the centre of the stellar sodium line, the signal of the transit would be less than one photon. In this scenario, we are not photon-noise limited anymore, but the signal is of the order of the detector red noise \citep{Pepe2021}. We have therefore masked the low-SNR remnant in the SRF for the FWHM of the stellar sodium lines (masking $3$ and $5\, \kms$ around line centre for night 1 and night 2, respectively). This reduces the number of spectra with planetary signal by $63$ ($39\%$) in night one and $88$ ($54\%$) in night two.

\subsection{Data quality assessment}
\label{sec:bootstrap}

\begin{figure*}
\begin{minipage}{0.49\linewidth}
\centering
\includegraphics[width=\textwidth]{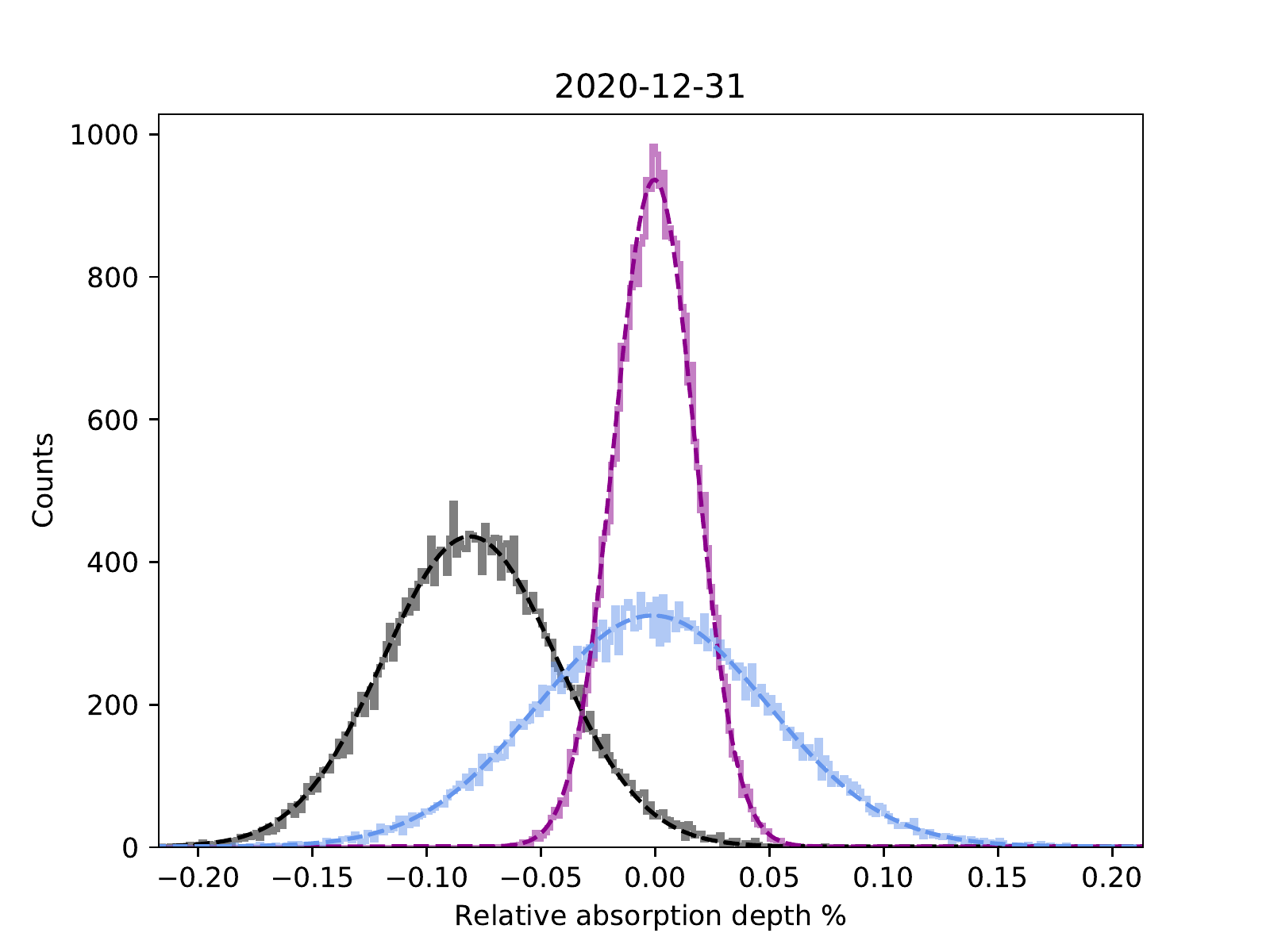}
\end{minipage}%
\begin{minipage}{0.49\linewidth}
\centering
\includegraphics[width=\textwidth]{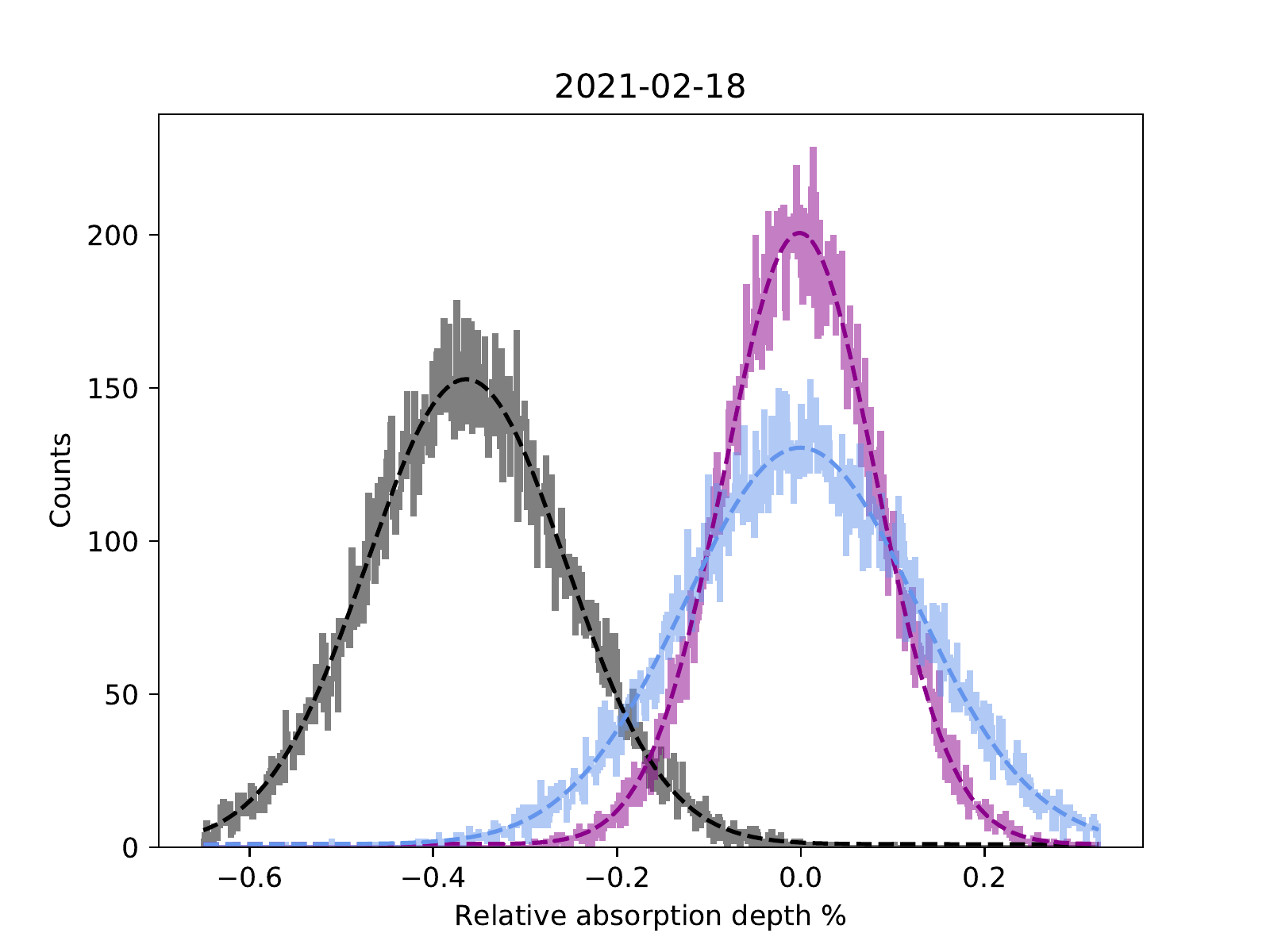}
\end{minipage}
\caption{Result of the EMC bootstrap analysis for the $12$ \r{A} passband for the two ESPRESSO transits. Night one is shown on the left and night two on the right. Both the `in-in' (purple) and `out-out' (light blue) distributions show no sodium detection, but the `in-out' distribution shows a detection (gray) for each night. Gaussian fits are shown as dashed lines. Because a different number of spectra were observed each night, the number of counts for the different randomisation varies.}
\label{fig:random}
\end{figure*}

\noindent To calculate the probability that a detection is a false-positive, due to instrumental effects, stellar spots or observational conditions varying during the night, we performed a bootstrap analysis via empirical Monte-Carlo (EMC). The false-positive probability is then taken into account in the calculation of the detection error. Especially for noisy observations, it is important to take into account this additional uncertainty, because the likelihood of a false-positive increases. During the bootstrap analysis the in- and out-of-transit spectra are treated like independent data sets and both the virtual in-transit and virtual out-of-transit spectra are randomly drawn from one data set only, following \citet{Re08} with the same applied wavelength regions. Three scenarios are created, the `in-in' (all spectra taken from the real in-transit data and randomly attributed to virtual in and out-of-transit data), the `out-out' (drawn only from out-of-transit spectra), and the `in-out' \citep[likewise; for further details see][]{Wy15,Seidel2019}. If only the scenario `in-out' results in a detection of any sort, the detection is unlikely to stem from spurious events.

The false positive likelihood is the standard deviation of the Gaussian fit to the `out-out' distribution, given that no planetary signal can influence the distribution, but within the standard deviation, a false detection is possible. Because of the selection bias when taking only the out-of-transit distributions, the final false positive likelihood has to be scaled by the square-root of the fraction of out-of-transit spectra to total spectra taken \citep{Re08,As13}.

\noindent The bootstrapping distributions for both nights are shown in Figure \ref{fig:random}, with $30,000$ iterations each for conversion of the distribution. For both nights, the `in-in' and `out-out' distributions are centred at $0$, which means no spurious features were detected. The `in-out' distribution shows the absorption depth unambiguously not equal to zero, showing the planetary origin on the signal. The false positive likelihood for night one is $0.023\,\%$ and for night two $0.070\,\%$

\subsection{Atmospheric absorption depth}
\label{sec:transspec}

 \begin{table}
\caption{Sodium detection levels.}
\label{table:detections}
\centering
\begin{tabular}{c | c | c c }
\hline
\hline
& & Detection [$\%$] & $\sigma$     \\
\hline
Night 1 & D$_2$ & $0.498\pm0.082$ & $6.0$\\
& D$_1$ & $0.349\pm0.081$ & $4.3$ \\
& D$_2$ and D$_1$ & $0.424\pm0.058$ & $7.3$ \\
\hline
Night 2& D$_2$ & $0.389\pm0.085$ & $4.6$\\
& D$_1$ & $0.275\pm0.086$ & $3.2$ \\
& D$_2$ and D$_1$ & $0.332\pm0.060$ & $5.5$ \\
\hline
\end{tabular}
\end{table}
The sodium doublet transmission spectra for \Planet{} is shown in full for each night in Figure \ref{fig:transspectrum} (in the PRF). The two peaks of the doublet are clearly visible and the line centres each highlighted as dashed, blue vertical lines. Both lines were jointly fitted with Gaussians, which are shown as red lines. The fitted amplitude of the Gaussians was used as the absorption depth and the uncertainty on the absorption as the sum of the false positive likelihood (calculated in Section \ref{sec:bootstrap}), the uncertainty of the Gaussian fit, and the average uncertainty within the FWHM of the sodium doublets \citep[see][]{Hoeijmakers2020}.

In both nights, a large fraction of the planetary signal was masked due to the low-SNR remnant (see Section \ref{sec:stellar}), with a larger masking applied in night two. The lower SNR in night two is most likely a result from the increase in seeing during the second half of the transit and the after transit baseline by a factor of two. However, the detection levels differ for both nights and the doublet is not as well resolved in night two as it is in night one, which cannot be solely explained by the broader masking in night two. Possible impact factors in night two are the small spead in barycentric velocity, leading to a larger overlap between possible micro-telluric residuals and the sodium doublet as well as instrumental issues (ESPRESSO was used on UT4, which was not used as habitually at the time). The detection levels split for both nights and both lines can be found in Table \ref{table:detections}. We chose to still present both datasets and hope that the exact line shape of \Planet{} can be resolved without any doubts with an additional observing night discriminating between the here presented night one and two.

\section{Interpretation of the obtained sodium feature}
\label{sec:interp}

\begin{figure}
\resizebox{\columnwidth}{!}{\includegraphics[trim=-0.0cm 1.5cm -0.0cm 2.0cm]{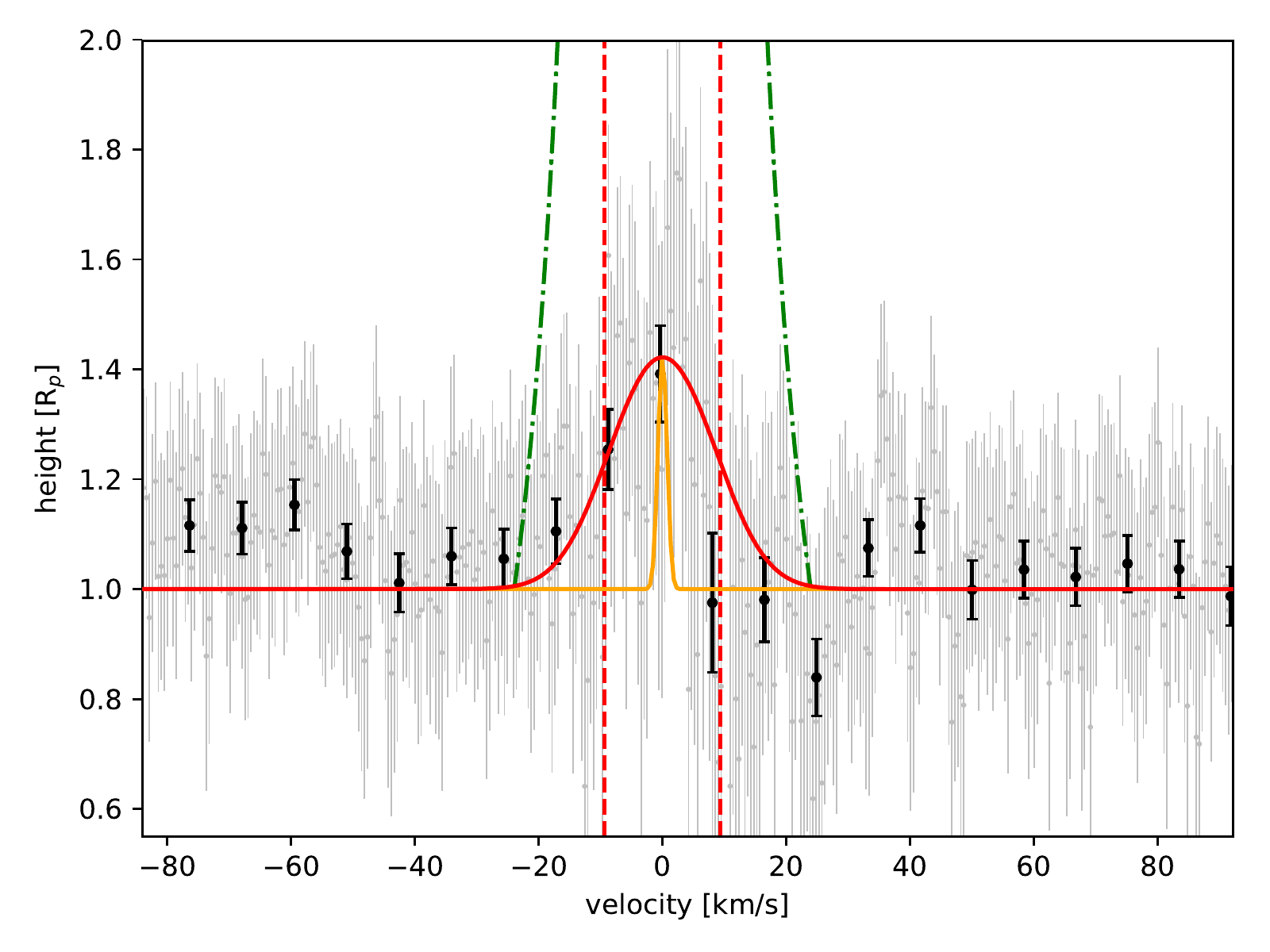}}
        \caption{Both sodium lines combined in velocity space as a function of altitude above the surface. The green, dashed-dotted lines indicate the escape velocity assuming the continuum is the planet surface. The yellow Gaussian indicates the line spread function of ESPRESSO. The red line is a Gaussian fit to the combined lines, masking outliers in the combined line shape in the red wing of the absorption feature. The red, dashed lines indicate the FWHM of the Gaussian fit.}
        \label{fig:combined_esc}
\end{figure}

To give a tentative interpretation of the data, we combined both nights and both lines in a coadded sodium doublet.
The FWHM of the fit to the coadded sodium doublet is $18.74\pm1.90\,\kms$, corresponding to a broadening of $9.37\pm0.95\,\kms$ in both directions from line centre. Comparing the ESPRESSO line spread function (shown in yellow in Figure \ref{fig:combined_esc}) indicates significant line broadening. However, the escape velocity of \Planet{} at the surface is $23.83\pm0.81\,\kms$, which means that the sodium, although accelerated to high velocities, is not escaping the planet's gravity assuming the sodium doublet probes down to the surface. This result is below the initial estimation of line broadening as provided in the HARPS analysis from \cite{Seidel2020b}. However, considering the significant uncertainties in the HARPS data, the results herein are not incompatible with the previous study. 

Due to the nature of transmission spectroscopy, the knowledge of the continuum level is lost. The assumption that the continuum is consistent with the planetary surface therefore might not hold if the atmosphere is cloudy \citep[e.g.][]{Morley2013, Lavvas2019}. We can, therefore, not assess with certainty if sodium is escaping the atmosphere of this significantly bloated hot Neptune without the study and modelling of water vapour in the red optical wavelength, which is out of the scope of this paper. As of now, no HST observations are published, however \Planet{} is scheduled to be observed with the newly launched JWST which should provide more insight into this intriguing target \citep{Mayo2021}.

\section{Conclusion}
\label{sec:conclusion}

We analysed two transits of \Planet{} observed with ESPRESSO, resulting in 570 exposures with 100 s exposure time each. Of these, thirteen spectra of the out-of-transit baseline in night one were rejected, resulting in 322 in-transit exposures and 220 out-of-transit exposures. Due to the significant stellar sodium absorption, the resulting deep stellar sodium lines, and the comparatively short exposure time, low-SNR remnants were produced. We masked these low-SNR zones and subsequently confirmed the HARPS sodium detection at the $7.3\,\sigma$ level and $5.5\,\sigma$ level respectively in each night, the first resolved sodium detection for a planet within the Neptune desert achieved by two independent instruments.

We suggest further analysis of WASP-166b not only with space-based facilities like JWST, but also with ground based facilities to establish the true line shape of the sodium doublet. Due to the low-SNR remnants, we loose a significant part of the planetary signal for each transit. We hope rapid follow-up observations of this target will give us a clearer picture of the line shape, allowing us to gain a first observational constraint on the winds in the upper atmosphere of a bloated hot Neptune in the hot Neptune desert \citep[i.e. ][]{Seidel2020a,Wyttenbach2020,Seidel2021}.


\section*{Acknowledgements}
This publication makes use of The Data \& Analysis Center for Exoplanets (DACE), University of Geneva (CH). DACE is a platform of PlanetS.
This work has received funding from a UKRI Future Leader Fellowship (grant number MR/S035214/1). R. A. is a Trottier Postdoctoral Fellow and acknowledges support from the Trottier Family Foundation. This work was supported in part through a grant from FRQNT.
\section*{Data availability}
The data is available from the ESO archive under programme ID 106.21EM.

%

\bibliographystyle{mnras} 
\bibliography{WASP166b_SODIUM} 

\bsp	
\label{lastpage}
\end{document}